\documentclass[a4paper,11pt]{article}

\usepackage{jheppub} 

\usepackage[T1]{fontenc} 

\usepackage{amsmath}

\title{\boldmath An Improvement for
Quantum Tunneling Radiation of Fermions
 in a Stationary Kerr-Newman Black Hole Spacetime}

\author[a,1]{J. Zhang,\note{Corresponding author.}}
\author[a]{M. Q. Liu,}
\author[a]{Z. E. Liu}
\author[b]{and S. Z. Yang}

\affiliation[a]{ College of Physics and Electronic Engineering, Qilu
Normal University\\2 Wenbo Road,  Jinan  250300,  China}
\affiliation[b]{College of Physics and Space Science, China West
Normal University\\1 Shida Road, Nanchong 637000 China}

\emailAdd{zhangjie\_mail@cqu.edu.cn}

\abstract{By introducing a specific etheric-like vector in the Dirac
equation with
  Lorentz Invariance Violation (LIV) in the curved spacetime, an improved method for
quantum tunneling radiation of fermions is proposed. As an example,
 we apply this new method to a charged axisymmetric
Kerr-Newman black hole. Firstly, considering LIV theory, we derive a
modified dynamical equation of fermion with spin 1/2 in the
Kerr-Newman black hole spacetime. Then we solve the equation and
find the increase or decrease of black hole's Hawking temperature
and entropy are related to constants $a$ and $c$ of the Dirac
equation with LIV in the curved spacetime. As $c$ is positive, the
new Hawking temperature is about $
\frac{\sqrt{1+2a+2cmk_r^2}}{\sqrt{1+2a}}$ times higher than that
without modification, but the entropy will decrease. We also make a
brief discussion for the case of high spin fermions.}

\begin{document}
\maketitle
\flushbottom

\section{Introduction}
Black hole is a mysterious cosmic body with extremely intense
gravity. With the detection of LIGO and Virgo, more and more
activities of black holes are found, so current theoretical
researches for black holes become more and more significant. Black
holes can be divided into three types: static black hole, stationary
black hole and dynamic black hole. Hawking firstly
 proved that black holes have thermal radiation in theory
by studying the quantum effect near the horizon of black holes
\cite{haw74,haw75}. Hawking radiation effectively links
gravitational theory, quantum theory and thermodynamic statistics
physics, and inspired other researchers to study the thermodynamical
evolution of black holes \cite{dam76,san88,zha91,yan95}. Parikh et
al. pointed that the behavior of Hawking radiation can be regarded
as a quantum tunneling \cite{par00,par06}. In their assumption, the
event horizon of black hole is a potential barrier, virtual
particles yielded inside the horizon have a certain probability to
escape from this barrier and  be converted to real particles
radiating out from the black hole. Refs.\cite{
liu07,akh06,sri99,sha02,med02,iso06,zha06,yan05,liu19} adopted the
quantum tunneling method to investigate Hawking radiation for
different types of black holes. Srinivasan et al. derived the
Hamilton-Jacobi equation in curved spacetime from a scalar field
equation 
\cite{sri99,sha02}.  Kerner and Mann et al. studied the tunneling
radiation of Dirac particles using a semi-classical theory
\cite{ker08a,ker08b,cri08}. Lin and Yang proposed a new method to
study the quantum tunneling radiation of fermions
\cite{lin09b,lin09a,lin11,yan10}. Their method can also be used to
study the quantum tunneling radiation of bosons. The results
obtained in Refs.\cite{lin09b,lin09a,lin11,yan10} show that the
Hamilton-Jacobi equation in curved spacetime is a basic equation of
particle dynamic, which reflects the inherent consistency between
Lorentz symmetry theory and the Hamilton-Jacobi equation. Recently,
we considered a light dispersion relationship derived from string
theory to research the modified quantum tunneling rates for
spherical symmetry and axisymmetryic black
holes\cite{zha20a,zha20b}.

 General relativity is a gravitational theory
that cannot be renormalized, so several modified gravitational
theories have been proposed. Since the Lorentz Invariation Violation
(LIV) may exist at high energy cases, various gravity models based
on LIV have been proposed \cite{hor09,jac01,lin14}. In principle,
LIV theory can solve the problem of renormalization of gravitational
theory. In addition, some studies on LIV suggest that the dark
matter may be just one of the effects of LIV theoretical models
\cite{muk10}. In string theory, electrodynamics and non-abel theory,
LIV has attracted extensive attention \cite{col07,jac99,kos89}. In
the recent years, Dirac equation with LIV term in the flat spacetime
has been studied by introducing etheric-like field terms
\cite{cas11,nas15}.  In this theory, Lorentz symmetry disappears due
to the existence of the ether-like field. Therefore, some special
properties which are inconsistent with Lorentz symmetry theory will
emerge at the high energy case. This is an interesting topic which
needs to be explored in depth. On the other hand, the dynamics of
fermions with LIV in the curved spacetime is also an attractive
subject worthy further study, which will influence the correction to
quantum tunneling radiation of black hole. At present, the quantum
tunneling radiation of Dirac particles with etheric-like field terms
has been investigated  in spherically symmetric black holes
\cite{pu19}. We investigate the influence of different etheric-like
vector $u^\alpha$ on the solution of the modified Hamilton-Jacobi
equation by using both the semi-classical approximation and beyond
the semi-classical approximation\cite{zha20c}.

 In this paper, the
quantum tunneling radiation of fermions is modified in the
axisymmetric charged Kerr-Newman black hole by considering a
specific ether-like field vector term. Our paper is organized as
follows: In Sec.2, considering LIV theory, the dynamical equation of
fermions with spin 1/2 is derived for Kerr-Newman black hole. In
Sec.3, we solve this dynamical equation and obtain the corrected
physical quantities such as Hawking temperature and tunneling rate
of the black hole. We make some discussions in Sec.4.

\section{Lorentz Invariance Violation Theory and Dirac-Hamilton-Jacobi Equation}

In the Ref.\cite{nas15}, Nascimento et al. researched the particle's
action which includes LIV in the flat spacetime. Transferring the
normal derivative to covariant derivative, and extending  the
commutation relation of gamma matrices $\bar{\gamma}^{\mu}$ and
$\bar{\gamma}^{\nu}$  in the flat spacetime to that in the curved
sapcetime, we obtain Dirac equation of fermion, spin of which is
1/2,  with LIV in the curved spacetime as:
\begin{equation} \label{eq:1}
\{\gamma^\mu D_{\mu}[1+\hbar^2\frac{a}{m^2}(\gamma^\mu
D_\mu)^2]+\frac{b}{\hbar}\gamma^5+c\hbar(u^\alpha
D_\alpha)^2-\frac{m}{\hbar})\}\Psi=0,
 \end{equation}
where $m$ is the mass of fermion, $a$, $b$ and $c$ is small
constants. Gamma matrices $\gamma^\mu$ satisfy the condition:
\begin{equation} \label{eq:2}
\gamma^{\mu}\gamma^{\nu}+\gamma^{\nu}\gamma^{\mu}=2g^{\mu \nu}I,
 \end{equation}
\begin{equation} \label{eq:3}
\gamma^{5}\gamma^{\mu}+\gamma^{\mu}\gamma^{5}=0,
 \end{equation}
 where $g^{\mu \nu}$ is the inverse metric tensor, and $I$  is the unit matrix.
 In the flat spacetime, Eq.(\ref{eq:2}) reduces to $\bar{\gamma}^{\mu}\bar{\gamma}^{\nu}+\bar{\gamma}^{\nu}\bar{\gamma}^{\mu}=2\delta^{\mu
 \nu}I$, and
 Eq.(\ref{eq:3}) changes to
 $\bar{\gamma}^{5}\bar{\gamma}^{\mu}+\bar{\gamma}^{\mu}\bar{\gamma}^{5}=0$.
 In Eq.(\ref{eq:1})
 \begin{equation} \label{eq:4}
D_{\mu}=\partial_{\mu}+\frac{i}{2}\Gamma^{\alpha \beta}_{\mu}
\pi_{\alpha \beta}-\frac{i}{\hbar} q A_{\mu},
 \end{equation}
where $q$ is the charge of fermion, and $A_\mu$ is the
electromagnetic potential of black hole. The second term at the
right side of Eq.(\ref{eq:4}) is spin connection which is a very
small term in the dynamical equation and thus can be ignored.
$u^\alpha$ is an etheric-like vector, which satisfies:
 \begin{equation} \label{eq:5}
u^\alpha u_\alpha={\rm const}.
 \end{equation}
 In order to solve the Dirac equation for fermions with spin 1/2, we
 assume its wave function is
 \begin{equation} \label{eq:6}
\Psi=\psi_{A B}{\rm e}^{\frac{{\rm i}}{\hbar}S}=\left(
\begin{array}{c}
A\\
B
\end{array}
\right ){\rm e}^{\frac{{\rm i}}{\hbar}S},
 \end{equation}

where $A$ and $B$ are matrix elements in the column matrix, $S$ is
the Hamilton principal function. Substituting Eq.(\ref{eq:4}) and
Eq.(\ref{eq:6}) into Eq.(\ref{eq:1}), we get
  \begin{eqnarray} \label{eq:7}
\{i \gamma^\mu(\partial
_{\mu}S-qA_\mu)[1-\frac{a}{m^2}\gamma^\alpha\gamma^\beta(\partial
_{\alpha}S-qA_\alpha)(\partial _{\beta}S-qA_\beta)]
\nonumber\\
-cu^\alpha u^\beta(\partial _{\alpha}S-qA_\alpha)(\partial
_{\beta}S-qA_\beta)+b\gamma^5-m\}\Psi=0.
   \end{eqnarray}
Using Eq.(\ref{eq:2}), we have
  \begin{equation} \label{eq:8}
\gamma^{\alpha}\gamma^{\beta}(\partial
_{\alpha}S-qA_\alpha)(\partial _{\beta}S-qA_\beta)=g^{\alpha
\beta}(\partial _{\alpha}S-qA_\alpha)(\partial _{\beta}S-qA_\beta).
   \end{equation}
Combining Eq.(\ref{eq:8}) and Eq.(\ref{eq:7}), they yields
 \begin{eqnarray}\label{eq:9}
i \gamma^\mu(\partial
_{\mu}S-qA_\mu)\Psi&=[1-\frac{a}{m^2}\gamma^\alpha\gamma^\beta(\partial
_{\alpha}S-qA_\alpha)(\partial _{\beta}S-qA_\beta)]^{-1}
\nonumber\\&\{cu^\alpha u^\beta(\partial
_{\alpha}S-qA_\alpha)(\partial
_{\beta}S-qA_\beta)+b\gamma^5+m\}\Psi \nonumber\\
&=[1+\frac{a}{m^2}\gamma^\alpha\gamma^\beta(\partial
_{\alpha}S-qA_\alpha)(\partial _{\beta}S-qA_\beta)+\mathcal{O}(a^2)]
\nonumber\\&\{cu^\alpha u^\beta(\partial
_{\alpha}S-qA_\alpha)(\partial _{\beta}S-qA_\beta)+b\gamma^5+m\}\Psi
\nonumber\\&=[1+(\frac{c}{m}u^\alpha
u^\beta+\frac{a}{m^2}g^{\alpha\beta})(\partial
_{\alpha}S-qA_\alpha)(\partial
_{\beta}S-qA_\beta)\nonumber\\&-\frac{b}{m}\gamma^5]m\Psi.
\end{eqnarray}
   Since $b\ll m$, so $\frac{b}{m}\gamma^5$ is very small. Multiplying $i\gamma^{\nu}(\partial_{\nu}S-qA_\nu)$ at both sides of Eq.(\ref{eq:9}), we get
\begin{eqnarray} \label{eq:10}
-\gamma^{\mu}\gamma^{\nu}(\partial _{\mu}S-qA_\mu)(\partial
_{\nu}S-qA_\nu)\Psi =
m^2+2(cmu^{\alpha}u^{\beta}+ag^{\alpha\beta})\nonumber\\(\partial
_{\alpha}S-qA_\alpha)(\partial _{\beta}S-qA_\beta)\Psi+\mathcal{O},
\end{eqnarray}
i. e.,
\begin{eqnarray} \label{eq:11}
[g^{\mu\nu}(\partial _{\mu}S-qA_\mu)(\partial
_{\nu}S-qA_\nu)\nonumber\\+2(cmu^{\mu}u^{\nu}+ag^{\mu\nu})(\partial
_{\mu}S-qA_\mu)(\partial _{\nu}S-qA_\nu)+m^2]\Psi = 0.
\end{eqnarray}
This is a matrix equation. The condition for this matrix equation to
have nontrivial solutions is that the value of determinant of the
wave function's coefficient matrix is zero, i. e.,
\begin{equation}\label{eq:12}
(g^{\mu\nu}+2cmu^{\mu}u^{\nu}+2ag^{\mu\nu})(\partial
_{\mu}S-qA_\mu)(\partial _{\nu}S-qA_\nu)+m^2 = 0,
\end{equation}
or
\begin{equation}\label{eq:13}
(g^{\mu\nu}+\frac{2cmu^{\mu}u^{\nu}}{1+2a})(\partial
_{\mu}S-qA_\mu)(\partial _{\nu}S-qA_\nu)+m^2/(1+2a) = 0.
\end{equation}
Considering $a$ is a very small constant and adopting Taylor
expansion for the last term in the Eq.(\ref{eq:13}), we get
\begin{equation}\label{eq:14}
(g^{\mu\nu}+\frac{2cmu^{\mu}u^{\nu}}{1+2a})(\partial
_{\mu}S-qA_\mu)(\partial _{\nu}S-qA_\nu)+m^2(1-2a) = 0.
\end{equation}
From Eq.(\ref{eq:1}) to Eq.(\ref{eq:14}), we get a new dynamical
equation for Dirac particles. Comparing with the normal
Hamilton-Jacobi equation, one can find that, as
$2cmu^{\mu}u^{\nu}=a=0$, Eq.(\ref{eq:14}) returns to the normal
Hamilton-Jacobi equation. We call this deformed equation as
Dirac-Hamilton-Jacobi equation. The choice of the Hamilton principal
function $S$ depends on the selected coordinate and line element. In
a stationary spacetime, it can be generally expressed as
$S=S(t,r,\theta,\varphi)$.

\section{Correction to Tunneling radiation of Fermions in the Kerr-Newman Spacetime}
In the Boyer-Lindquist coordinate, the line element of Kerr-Newman
black hole is written as \cite{wal84}
\begin{equation} \label{eq:15}
ds^2
=\rho^2(\frac{dr^2}{\Delta}+d\theta^2)+\frac{\sin^2\theta}{\rho^2}[(r^2+a_{kn}^2)d\varphi-a_{kn}dt]^2-\frac{\Delta}{\rho^2}[dt-a_{kn}\sin^2\theta
d\varphi]^2,
 \end{equation}
where
\begin{eqnarray} \label{eq:16}
&\Delta =r^2+a_{kn}^2-2Mr+Q^2, \nonumber\\
 &\rho^2= r^2+a_{kn}^2\cos^2\theta,
 \end{eqnarray}
where $M$ and $a$ are the mass and angular momentum of unit mass of
black hole. From Eq.(\ref{eq:15}) and Eq.(\ref{eq:16}), one can get
the components of non-zero covariant metric tensors
\begin{eqnarray} \label{eq:17a}
&g_{tt}=\frac{1}{\rho^2}(a_{kn}^2\sin^2\theta-\Delta),\nonumber\\
&g_{rr}=\frac{\rho^2}{\Delta},\qquad\qquad\nonumber\\
 &g_{\theta\theta}=\rho^2,\qquad\qquad\qquad\quad\nonumber\\
 &g_{\varphi\varphi}=\frac{\sin^2\theta}{\rho^2}[(r^2+a_{kn}^2)-\Delta a_{kn}^2\sin^2\theta],\nonumber\\
 &g_{t\varphi}=\frac{a_{kn}\sin^2\theta}{\rho^2}(Q^2-2Mr),\,
 \end{eqnarray}
  and inverse metric tensors
  \begin{eqnarray} \label{eq:17b}
&g^{tt}=\frac{1}{\rho^2}(a_{kn}^2\sin^2\theta-\frac{r^2+a_{kn}^2}{\Delta}),\nonumber\\
 &g^{rr}=\frac{\Delta}{\rho^2},\nonumber\\
 & g^{\theta\theta}=\frac{1}{\rho^2},\nonumber\\
 & g^{\varphi\varphi}=\frac{1}{\rho^2}(\frac{1}{\sin^2\theta}-\frac{a_{kn}^2}{\Delta}),\nonumber\\
 & g^{t\varphi}=\frac{1}{\rho^2}(\frac{Q^2-2Mr}{\Delta}).
 \end{eqnarray}
According to null super-surface equation
\begin{equation}
\label{eq:18}
 g^{\mu\nu}\frac{\partial f}{\partial
x^{\mu}}\frac{\partial f}{\partial x^{\nu}}=0,
 \end{equation}
the event horizon of black hole $r_H$ satisfies following equation
\begin{equation} \label{eq:19}
\Delta|_{r=r_H}=r^2_H-2Mr_H+a_{kn}^2+Q^2=0.
 \end{equation}
The electromagnetic potential  of the Kerr-Newman black hole
\begin{eqnarray} \label{eq:20}
A_\mu&=(A_t,0,0,A_\varphi), \nonumber\\
&A_t=-\frac{Qr}{\rho^2},\nonumber\\
&A_\varphi=\frac{a_{kn}Qr\sin^2\theta}{\rho^2}.
 \end{eqnarray}
 Substituting Eq.(\ref{eq:17a}) and (\ref{eq:17b}) into Eq.(\ref{eq:14}),  we get the
 dynamical equation of spin 1/2 fermions with mass $m$ and charge $q$ in the curved
 spacetime,
  \begin{eqnarray} \label{eq:21}
&g^{tt}(\partial_{t}S-qA_{t})^2+2g^{t\varphi}(\partial_{t}S-qA_{t})(\partial_{\varphi}S-qA_{\varphi})+g^{rr}(\partial_{r}S)^2+g^{\theta\theta}(\partial_{\theta}S)^2\nonumber\\
&+g^{\varphi\varphi}(\partial_{\varphi}S-qA_{\varphi})^2+(1-2a)m^2+\frac{2mc}{1+2a}[u^tu^t(\partial_{t}S-qA_{t})^2+u^ru^r(\partial_{r}S)^2\nonumber\\
&+u^\theta u^\theta(\partial_{\theta}S)^2+u^\varphi
u^\varphi(\partial_{\varphi}S-qA_{\varphi})^2+2u^t
u^\varphi(\partial_{t}S-qA_{t})(\partial_{\varphi}S-qA_{\varphi})\nonumber\\
&+2u^t u^r(\partial_{t}S-qA_{t})\partial_{r}S+2u^t
u^\theta(\partial_{t}S-qA_{t})\partial_{\theta}S\nonumber\\
&+2u^ru^\varphi\partial_{r}S(\partial_{\varphi}S-qA_{\varphi})+2u^\theta
u^\varphi\partial_{\theta}S(\partial_{\varphi}S-qA_{\varphi})]=0.
 \end{eqnarray}
After substituting the components of $g^{\mu \nu}$ into
Eq.(\ref{eq:21}), multiplying $\rho^2$ on both sides of the equation
and merging the similar terms, the dynamical equation are as
follows:
 \begin{eqnarray} \label{eq:22}
&\Delta(\frac{\partial S}{\partial
r})^2-\frac{1}{\Delta}[(r^2+a_{kn}^2)\frac{\partial S}{\partial t}
+a_{kn}\frac{\partial S}{\partial \varphi}+eQr]^2+r^2m^2(1-2a)+(\frac{\partial S}{\partial \theta})^2\nonumber\\
&+(\frac{1}{\sin\theta}\frac{\partial S}{\partial
\varphi}+a_{kn}\sin \theta\frac{\partial S}{\partial
t})^2+\rho^2m^2(1-2a)\cos^2\theta
+\frac{2mc \rho^2}{1+2a}[ u^t u^t(\frac{\partial S}{\partial t}-qA_t)^2 \nonumber\\
&+u^r u^r(\frac{\partial S}{\partial r})^2+ u^\theta
u^\theta(\frac{\partial S}{\partial \theta})^2+u^\varphi
u^\varphi(\frac{\partial
S}{\partial_\varphi}-qA_{\varphi})^2\nonumber\\&+2u^t
u^\varphi(\frac{\partial S}{\partial t}-qA_{t})(\frac{\partial
S}{\partial_\varphi}-qA_{\varphi})+2u^t u^r(\partial_{t}S-qA_{t})\partial_{r}S\nonumber\\
&+2u^t
u^\theta(\partial_{t}S-qA_{t})\partial_{\theta}S+2u^ru^\varphi\partial_{r}S(\partial_{\varphi}S-qA_{\varphi})\nonumber\\
&+2u^\theta
u^\varphi\partial_{\theta}S(\partial_{\varphi}S-qA_{\varphi})]=0.
 \end{eqnarray}
To solve the above equation, we must choose special $u^t$, $u^r$,
$u^\theta$ and $u^\varphi$ which must satisfy Eq.(\ref{eq:5}).
According to Eqs.(\ref{eq:15})-(\ref{eq:17b}) and the metric tensor
of Kerr-Newman black hole, we choose the  following $u^\alpha$:
\begin{eqnarray}\label{eq:23}
&u^t=\frac{k_t\rho}{(a_{kn}^2\sin^2\theta-\Delta)^{1/2}},\nonumber\\
&u^r=\frac{k_r \Delta^{1/2}}{\rho},\nonumber\\
&u^{\theta}=\frac{k_{\theta}}{\rho},\nonumber\\
&u^{\varphi}=\frac{k_{\varphi}\rho}{\sin\theta a_{kn}(Q^2-2Mr)},
\end{eqnarray}
where  $k_t,\ k_r,\ k_{\theta},\ k_{\varphi}$ are constants, which
satisfies
 \begin{equation} \label{eq:24}
u^\alpha u_\alpha={k^2_\alpha},
 \end{equation}
 where $\alpha$ denotes $t,\ r,\ \theta,\ \varphi$. Substituting Eq.(\ref{eq:23}) and Eq.(\ref{eq:19}) into Eq.(\ref{eq:22}),
and considering the limit at the event horizon of black hole, we can
get the dynamical equation of spin $1/2$ fermions at the event
horizon of black hole, that is,
\begin{equation}\label{eq:25}
\Delta^2\left|_{r\rightarrow r_H}\right.
(1+\frac{2cmk^2_r}{1+2a})\left(\frac{\partial S}{\partial
r}\right)^2\left|_{r\rightarrow r_H}\right.
-\left[(r_H^2+a_{kn}^2)\frac{\partial S}{\partial
t}+a_{kn}\frac{\partial S}{\partial \varphi}+qQr_H\right]^2 = 0.
\end{equation}
Because the quantum tunneling radiation of a black hole is a
property of the radial direction of the black hole, we care the
radial component of the Hamilton principal function. From
Eq.(\ref{eq:25}),
\begin{equation}\label{eq:26}
\frac{\partial S}{\partial r}\left|_{r \rightarrow
r_H}\right.=\pm\frac{(r_H^2+a_{kn}^2)\sqrt{1+2a}}{\Delta\left|_{r
\rightarrow r_H}\right. \sqrt{1+2a+2cmk^2_r}} \left(\frac{\partial
S}{\partial t}+\frac{a_{kn}\frac{\partial S}{\partial
\varphi}+qQr_H}{r_H^2+a_{kn}^2}\right).
\end{equation}
According to Eq.(\ref{eq:15}), we set 
\begin{equation}\label{eq:27}
S=-\omega t+R(r)+\Theta(\theta)+j\varphi,
\end{equation}
where $\omega$ is particle energy, $j$ is a constant which describes
the $\varphi$ component of general momentum. Eq.(\ref{eq:26}) can be
reducible to
\begin{equation} \label{eq:28}
\frac{\partial S}{\partial r}\left|_{r \rightarrow
r_H}\right.=\frac{dR}{dr}\left|_{r \rightarrow r_H}\right.
=\pm\frac{(r_H^2+a_{kn}^2)\sqrt{1+2a}}{\Delta\left|_{r \rightarrow
r_H}\right. \sqrt{1+2a+2cmk^2_r}} (\omega-\omega_0),
\end{equation}
where
\begin{equation} \label{eq:29}
\omega_0=\frac{qQr_H+a_{kn}j}{r_H^2+a_{kn}^2},
\end{equation}
where $\omega$ is the particle energy, $\omega_0$ is the chemical
potential, which means the minimal energy of emission particles.
Integrating the above equation from the inner side to the outer side
of $r_H$ with the residue theorem, we obtain
\begin{eqnarray}\label{eq:30}
S_{\pm}=R_{\pm}&=\pm\int dr \frac{(r_H^2+a_{kn}^2)\sqrt{1+2a}}{\Delta\left|_{r \rightarrow r_H}\right. \sqrt{1+2a+2cmk^2_r}}(\omega-\omega_0),\nonumber\\
&=\pm \frac{i \pi}{2} \frac{(r_H^2+a_{kn}^2)\sqrt{1+2a}}{(r_H-M)
\sqrt{1+2a+2cmk^2_r}}(\omega-\omega_0),
\end{eqnarray}
where subscript + and - denote outgoing and incoming module,
respectively. According to the tunneling theory of black hole, the
tunneling rate is
 \begin{eqnarray} \label{eq:31}
\Gamma &=\exp[-2({\rm Im} S_+-{\rm Im} S_{-}) ]\nonumber\\
&=\exp(-\frac{\omega-\omega_0}{T^{'}_H}),
 \end{eqnarray}
where $T^{'}_H$ is Hawking temperature after modification.

\begin{eqnarray} \label{eq:32}
T^{'}_H&=\frac{(r_H-M)(1+2a+2cmk^2_r)^{1/2}}{2\pi
(r^2_H+a_{kn}^2)\sqrt{1+2a}}\nonumber\\
&=T_{KNh}(1+2a+2cmk^2_r)^{1/2}/\sqrt{1+2a}, \end{eqnarray}
 where $T_{KNh}$ is the
Hawking temperature without LIV modification,
\begin{equation} \label{eq:33}
T_{KNh}=\frac{(r_H-M)}{2\pi(r^2_H+a_{kn}^2)}.
 \end{equation}

 \section{Summery and Discussions}
  In this paper, we have modified the dynamical equation of Dirac
particles in the curved spacetime, considering LIV theory. Comparing
Eq.(1) and Eq.(\ref{eq:14}), we simplify the complicate derivation
process. By solving Eq.(\ref{eq:14}), the new Hamilton principal
function $S$ of Dirac particle is obtained. Since the quantum
tunneling rate and Hawking temperature at the horizon of black hole
depend on the imaginary part of $S$,  so the new quantities related
to Hawking radiation are obtained naturally. These results are
valuable for further study of LIV theory and quantum gravitational
theory. Because LIV theory modifies the Hawking temperature at the
event horizon of Kerr-Newman black hole, it will lead to the
correction of black hole entropy. According to the first law of
thermaldynamics of black hole,
\begin{equation} \label{eq:34}
{\rm d}M=T{\rm d}S +V{\rm d}J +U{\rm d}Q,
 \end{equation}
where $V$ and $U$ are the rotation potential and electric potential
of black hole respectively. From Eq.(\ref{eq:32}) and
Eq.(\ref{eq:33}), the entropy with LIV theory correction in the
Kerr-Newman black hole is
\begin{eqnarray}\label{eq:35}
S_H=\int d S_H&=\int \frac{dM-VdJ-UdQ}{T_H}\nonumber\\
&= \int \frac{\sqrt{1+2a}}{\sqrt{1+2a+2cmk_r^2}}dS_{KNh},
\end{eqnarray}
where $S_{KNh}$ denotes the entropy without modification,
\begin{equation}\label{eq:36}
dS_{KNh}=\frac{dM-VdJ-UdQ}{T_{KNh}}.
\end{equation}
It can be seen from Eq.(\ref{eq:32}) and Eq.(\ref{eq:35}) that the
increase or decrease of black hole's Hawking temperature and entropy
are related to the value of constants $a$ and $c$ in
Eq.(\ref{eq:1}). As $c$ is positive, the new Hawking temperature is
about $ \frac{\sqrt{1+2a+2cmk_r^2}}{\sqrt{1+2a}}$ times higher than
that without modification, but the entropy will decrease. Above
results are from semi-classical approximation and based on spin
$3/2$ fermions. Beyond semi-classical approximation can refer our
recent paper \cite{zha20c}. For spin $3/2$ fermions, the wave
function in Eq.(\ref{eq:1}) should be substituted by
\begin{equation}\label{eq:36}
\Psi=\left(
\begin{array}{c}
A_{V}\\
B_{V}
\end{array}
\right ){\rm e}^{\frac{{\rm i}}{\hbar}S},
\end{equation}
where $A_V=[A \ \ B]^T$, $B_V=[A \ \  B]$\cite{yan10}. For arbitrary
spin fermions, $\Psi$ in Eq.(\ref{eq:1}) should be substituted by
$\Psi_{\alpha_1\ldots\alpha_k}$, where the value of $\alpha_k$
corresponds to different spin. The larger $\alpha_k$, the higher the
spin is.  Furthermore, the above conclusions are also applicable to
the other spherically symmetric and axisymmetric charged black
holes. We will do further research at this aspect in the future.

\section*{Acknowledgments}
 This work is supported by the National Natural Science Foundation of
China (grant U2031121), the Science Foundation of Sichuan Science
and Technology Department(grant 2018JY0502) and the Natural Science
Foundation of Shandong Province (grant ZR2020MA063, ZR2019MA059).


\end{document}